\documentclass[twocolumn,showpacs,amsmath,amssymb,floatfix,superscriptaddress]{revtex4}
\usepackage{graphicx}
\usepackage{dcolumn}
\usepackage{bm}
\usepackage{hyperref} 
\usepackage{latexsym}

\begin{document}
\title{Majority versus minority dynamics: \\ Phase transition in an interacting
  two-state spin system}
\author{M. Mobilia and S.~Redner}
\email{{mmobilia,redner}@bu.edu}
\affiliation{
Center for BioDynamics,
Center for Polymer Studies \&  Department of Physics, Boston University,
Boston, MA, 02215 }

\begin{abstract}
  
  We introduce a simple model of opinion dynamics in which binary-state
  agents evolve due to the influence of agents in a local neighborhood.  In a
  single update step, a fixed-size group is defined and all agents in the
  group adopt the state of the local majority with probability $p$ or that of
  the local minority with probability $1-p$.  For group size $G=3$, there is
  a phase transition at $p_c=2/3$ in all spatial dimensions.  For $p>p_c$,
  the global majority quickly predominates, while for $p<p_c$, the system is
  driven to a mixed state in which the densities of agents in each state are
  equal.  For $p=p_c$, the average magnetization (the difference in the
  density of agents in the two states) is conserved and the system obeys
  classical voter model dynamics.  In one dimension and within a Kirkwood
  decoupling scheme, the final magnetization in a finite-length system has a
  non-trivial dependence on the initial magnetization for all $p\ne p_c$, in
  agreement with numerical results.  At $p_c$, the exact 2-spin correlation
  functions decay algebraically toward the value 1 and the system coarsens as
  in the classical voter model.

\end{abstract}

\maketitle 
\date{\today}

\section{Introduction}
 
\begin{figure}[ht] 
 \vspace*{0.cm}
 \includegraphics*[width=0.3\textwidth]{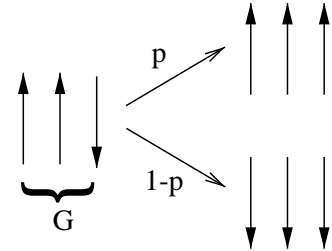}
\caption{Evolution of a group of $G=3$ spins according to MM dynamics.
  Majority rule applies with probability $p$ and minority rule applies with
  probability $1-p$.}
\label{update}
\end{figure}

In this article, we investigate the properties of a simple model of opinion
formation.  The model consists of $N$ agents, each of which can assume one of
two opinion states of $+1$ or $-1$.  These agents evolve according to the
following rules (Fig.~\ref{update}):

\begin{itemize}
  
\item Pick a group of $G$ agents (spins) from the system, with $G$ an odd
  number.  This group could be any $G$ spins in the mean-field limit, or it
  could be a randomly-chosen contiguous cluster of spins in
  finite-dimensional systems.
  
\item With probability $p$, the spins in the group all adopt the state of the
  local majority.  With probability $1-p$, the spins all adopt the state of
  the local minority.
  
\item Repeat the group selection and attendant spin update until the system
  necessarily reaches a final state of consensus.
\end{itemize}

We term this process the {\it majority-minority model\/} (MM), in keeping
with the feature that evolution can be controlled either by the local
majority or the local minority.  The MM model represents a natural outgrowth
of recent analytical work on the {\it majority rule} model of opinion
formation \cite{Majority}, which, in turn, represents a particular limit of a
class of models introduced by Galam \cite{galam1}.  In majority rule, the
opinion evolution of any group is controlled only by the local majority
within that group.  Thus majority rule corresponds to the $p=1$ limit of the
present MM model.

A basic motivation for this type of modeling is to incorporate, within a
minimalist description, some realistic aspects of the manner in which members
of an interactive population form consensus on some issue.  In this spirit,
the MM model allows for the possibility that a forceful and/or charismatic
minority can sometimes dominate the opinion of a group, an experience that
many of us have had in our everyday lives.  The limit where $p$ is close to 1
is probably closer to socially realistic situations.  Part of our interest in
consider the case of general $p$ is to understand the change in dynamics as a
function of $p$ and the kinetic phase transition that occurs at $p_c$.

We shall see that the interplay between minority and majority rule leads to
three distinct kinetic phases in which the approach to ultimate consensus is
governed by different mechanisms.  As in the earlier work on majority rule
\cite{Majority}, we seek to understand the long-time opinion evolution.  We
will be primarily concerned with determining the probability of reaching a
given final state (the exit probability) as a function of $p$ and the initial
densities of each opinion state.
  
To provide perspective for this paper, we briefly review related work on
opinion dynamics models.  Perhaps the simplest such example in this spirit is
the classical voter model \cite{voter}.  Here a 2-state spin is selected at
random and it adopts the opinion of a randomly-chosen neighbor.  This step is
repeated until a finite system necessarily reaches consensus.  One can think
of each spin as an agent with zero self confidence who merely adopts the
state of one of its neighbors.

An attractive feature of the voter model (in contrast to the familiar Ising
model with Glauber kinetics \cite{glauber}) is that it is exactly soluble in
all spatial dimensions.  For a finite system of $N$ spins in $d$ dimensions,
the time to reach consensus scales as $N$ for $d>2$, as $N\ln N$ for $d=2$
(the critical dimension of the voter model), and as $N^2$ in $d=1$
\cite{voter,cox,pk}.  In $d=1$ and 2, an infinite system coarsens so that
consensus emerges on progressively larger length scales, while for $d>2$, an
infinite system approaches a steady state of mixed opinions. Because the
average magnetization is conserved \cite{voter}, the probability that the
system eventually ends with all plus spins equals the initial density of plus
spins in all spatial dimensions.

From a more practically-minded viewpoint, there has been a recent upsurge of
interest in kinetic spin-based statistical physics models that attempt to
incorporate some realistic sociological features.  One such example is
Galam's rumor formation model \cite{galam1,galam2}, in which a population is
partitioned into variable-sized groups, and in each update step the spins in
each group may adopt the majority state or the minority state of the group
depending on additional interactions.  Our majority model represents a
special case in which only a single group of fixed size $G$ is updated at
each step.  Another prominent example is the Sznajd model, where spins evolve
only when local regions of consensus exist \cite{Sz}.  In the basic version
of the model, when two neighboring spins are in the same state, this local
consensus persuades a neighboring spin to join in.  Such a rule naturally
leads to eventual global consensus except in the anomalous case of an
antiferromagnetic initial state.  The generic questions posed above about
opinion evolution in the MM model are also of basic interest in the Sznajd
model \cite{SL03} and considerable work has recently appeared to quantify its
basic properties \cite{SL03,other,other1,other2,other3}.  There is also a
wide variety of kinetic spin models of social interactions that incorporate,
for example, multiple traits \cite{Axel}, incompatibility \cite{VKR,BKR}, and
other relevant features \cite{general}.

A new feature of our MM model is that the competition between majority and
minority rule leads to a kinetic phase transition in all spatial dimensions
$d$ at a critical value of $p_c= 2/3$ for group size $G=3$.  The existence of
such a transition can be easily understood by considering the average change
of the magnetization in a single update step.  A group undergoing an update
must consist of 2 spins of one sign and a single spin of the opposite sign.
According to Fig.~\ref{update}, the magnetization change in such a group is
proportional to $2p-4(1-p)$, which is zero when $p=p_c=2/3$.  For $p>p_c$ and
for all $d\geq 2$, the system quickly evolves toward global consensus where
the magnetization equals $\pm 1$ \cite{unfinished}.  For $p=p_c$, the average
magnetization is conserved, as in the voter model.  Consensus is again always
reached, but the time until consensus grows as a power law in time.  For
$p<p_c$, the system is driven toward a state with equal densities for the two
species of agents.  Since consensus is still the only absorbing state of the
dynamics, consensus is eventually reached in a finite system, but the time
needed grows exponentially with the system size.  It bears emphasizing that
for all $p$ and for all $d$, a finite-size system necessarily reaches
consensus in the MM model.  There are no metastable states that prevent the
attainment of ultimate consensus as in the related majority vote process
\cite{voter} or in the zero-temperature Ising Model with Glauber kinetics
\cite{SKR}.

The MM model exhibits special behavior in one dimension in which the
magnetization quickly approaches a static value that depends only on the
initial magnetization.  If one focuses on the interfaces between domains of
agents in the same state, these domain wall particles undergo the diffusive
annihilation reaction $A+A\to 0$, but with constraints in the motion of
domain walls, when they are nearby, that reflect the constraints of the MM
dynamical rules.  Our understanding of this intriguing aspect of the problem
is still incomplete.

In Sec.~II, we investigate the exit probability and exit times in the
mean-field limit of the MM model.  We then turn to the case of one dimension
in Sec.~III.  We first write the master equation for the configurational
probability distribution, following the original Glauber formalism.  We apply
a Kirkwood decoupling scheme \cite{K} for correlation functions to compute
the final magnetization as a function of the initial magnetization.  Finally,
we show that in the exactly solvable case of $p=p_c=\frac{2}{3}$, the 2-spin
correlation function $c_r (t) \equiv \langle S_i (t) S_{i+r} (t)\rangle$
approaches one as $t^{-1/2}$ for all $r$.  Thus the system exhibits diffusive
coarsening, as in the traditional voter model.  We give a summary and
discussion in Sec.~IV.  Calculational details are given in the appendices.

\section{The Mean-Field Limit}

\subsection{Exit Probability}

Following the approach developed in \cite{Majority}, we first study the exit
probability $E_n$, namely, the probability that a system that initially
contains $n$ up spins in a system of $N$ total spins ends with all spins up.
This exit probability obeys a simple recursion relation in which $E_n$ can be
expressed in terms of the exit probabilities after one step of the MM process
\cite{fpp}.

To construct this recursion relation, we note that
\begin{eqnarray*}
p_n\equiv 3p{N-3\choose n-2}\Big/{N\choose n}\quad {\rm and} \quad
q_n\equiv 3p{N-3\choose n-1}\Big/{N\choose n}
\end{eqnarray*}
are the respective probabilities that a group of 3 spins contains 2 plus and
1 minus spins or contains 1 plus and 2 minus spins, and that the majority
rule is applied to the group.  Thus $p_n$ is the probability that there is a
change $n\to n+1$ and $q_n$ is the probability that there is a change $n\to
n-1$ in a single step of the MM process.  Similarly
\begin{eqnarray*}
\overline p_n\equiv 3q{N-3\choose n-1}\Big/{N\choose n}\quad {\rm and}\quad
\overline q_n\equiv 3q{N-3\choose n-2}\Big/{N\choose n},
\end{eqnarray*}
with $q=1-p$, are the respective probabilities for $n$ to change by $\pm 2$
steps due to minority rule being applied to the group.  The master equation
for the exit probability is \cite{fpp}
\begin{equation}
\label{ME-exit}
E_n=\overline p_nE_{n+2}+ p_nE_{n+1}+q_nE_{n-1}+\overline q_nE_{n-2}
\end{equation}

\begin{figure}[ht] 
 \vspace*{0.cm}
 \includegraphics*[width=0.35\textwidth]{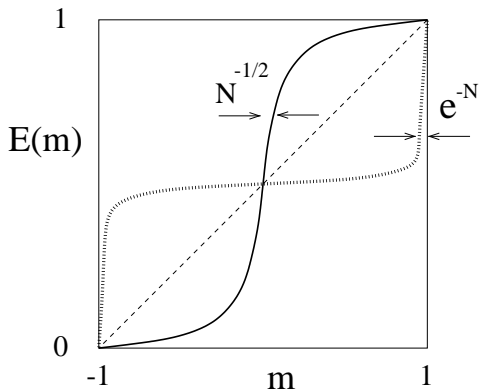}
\caption{Sketch of the exit probability $E(m)$ that a finite system with 
  initial magnetization $m$ ends with all spins plus for $p>p_c=2/3$ (solid),
  $p=p_c$ (dashed), and $p<p_c$ (dotted).  Also indicated is the $N$
  dependence of the deviation of the first and last curves from a step
  function.}
\label{E-sketch}
\end{figure}

While the exact solution to this discrete recursion relation was given in
\cite{Majority} (for $p=1$), it is much simpler to consider the 
continuum limit of
$n,N\to\infty$ with $x=n/N$ finite.  In this limit, the hopping probabilities
reduce to
\begin{eqnarray*}
p_n&=&3px^2(1-x), \quad q_n=3px(1-x)^2, \nonumber\\
\overline p_n&=&3qx(1-x)^2, \quad \overline q_n=3qx^2(1-x),
\end{eqnarray*}
and after some straightforward steps, the continuum version of the master
equation simplifies to
\begin{equation}
\label{ME-cont}
(3p-2)\,Nm\,E'(m)+(4-3p)\,E''(m)=0,
\end{equation}
where $m=2x-1$ is the magnetization and the prime denotes differentiation with
respect to $m$.  This equation can be easily integrated and the final result
is
\begin{equation}
\label{Em}
E(m)=\frac{1}{2}\left( 1 + \frac{I(m)}{I(1)}\right),
\end{equation}
where 
\begin{equation}
I(m)=\int_0^{\sqrt{m}}\,e^{-N\alpha y^2/2}\, dy, \nonumber
\end{equation}
with $\alpha=(3p-2)/(4-3p)$.

The behavior of $E(m)$ versus $m$ is sketched in Fig.~\ref{E-sketch} and it
merely represents the continuum version of the corresponding result given in
Ref.~\cite{Majority}.  For $p>p_c$, the exit probability approaches a step
function as $N\to\infty$ with a characteristic width that scales as
$N^{-1/2}$.  This feature reflects the fact that when $|m|>N^{-1/2}$, the
hopping process underlying the exit probability is controlled by the global
bias.  Conversely, for $p<p_c$, the exit probability approaches 1/2 for
nearly all initial values of $m$ except for a thin region of width $e^{-N}$
about $m=\pm 1$.  This reflects the fact that minority rule tends to drive
the system toward zero magnetization.  Thus the exit probability is
independent of the initial state unless the system begins at an exponentially
small distance (in $N$) from consensus.

\subsection{Magnetization}

The average magnetization also obeys a simple rate equation in the continuum
limit.  With probability $3x^2(1-x)$, where $x=n/N$, a group of 3 consists of
2 plus spins and 1 minus spin.  If this group is picked, then majority rule
applies with probability $p$ and the magnetization increases by 2, while with
probability $q$, the magnetization decreases by 4.  A complementary reasoning
applies to a group with 2 minus spins and 1 plus spin.  Thus the rate
equation for the magnetization is
\begin{eqnarray}
\label{magnMF}
\frac{dm}{dt}  &=& 6 x^2(1-x)(p-2q) - 6 x (1-x)^2 (p-2q)\nonumber \\
&=& 6(3p-2)\,m(1 -m^{2}),
\end{eqnarray}
where again $m=2x-1$.  This approximate equation becomes an exact description
in the limit $N\to\infty$.  The long-time solution is
\begin{equation}
\label{MFmagn}
m(t)\simeq \begin{cases}
{\displaystyle \pm\left\{1-\left[\frac{1-m^{2}(0)}{m^2(0)}\right]
e^{-36(p-p_c)t}\right\}}&
{\displaystyle p>p_c},\\ \\
{\displaystyle m(0)} & {\displaystyle p=p_c},\\ \\
{\displaystyle \frac{m(0)}{\sqrt{1-m^{2}(0)}}\, e^{-18(p_c-p)t}} & {\displaystyle p<p_c},
\end{cases}
\end{equation}
where in the first line, the $\pm$ sign occurs if $m(0)> 0$ or $m(0)<0$
respectively.

For $p>p_c$, majority rule prevails and the dynamics is essentially the same
as in the original majority rule model \cite{Majority}.  The approach to the
asymptotic behavior is exponential in time with a relaxation time $\tau_M=
[36(p-p_c)]^{-1}$.  This corresponds to an exit time that scales
logarithmically in the system size.  Conversely, when $p<p_c$ the dynamics is
dominated by the rule of the minority so that the asymptotic magnetization
vanishes (for $m(0)\neq \pm 1$).  The approach towards this steady state is
again exponential, but with a relaxation time $\tau_m=[18(p_c-p)]^{-1}$ that
is twice as large as $\tau_M$.  In spite of the bias away from consensus,
this state is necessarily reached in a finite system, because this is the
only absorbing state of the dynamics, but the time required to reach
consensus grows exponentially in the system size.  Finally, at the critical
point $p_c=2/3$, the average magnetization remains invariant, as in the voter
model \cite{voter}.

\section{MM MODEL IN ONE DIMENSION}

\subsection{Equations of Motion}

In one dimension, the original formalism of the Ising-Glauber model
\cite{glauber} can be exploited to obtain the equation of motion for the
magnetization, as well as that for higher-order spin correlation functions.
We consider only the simplest case of group size equal to three and denote
the spins in a group, that can take the values $\pm 1$, by $S$, $S'$ and
$S''$.  Then the rate at which spin $S$ flips according to majority rule is
\cite{Majority},
\begin{eqnarray}
\label{sfMM}
W(S\rightarrow -S)=1+S'S''-S(S'+S'').
\end{eqnarray}
This rate expresses the fact that $S'$ and $S''$ must be equal but opposite
to $S$ for spin $S$ to flip.  Conveniently, this same expression also gives
the rate at which the spins $S'$ and $S''$ flip according to minority rule
dynamics.  Thus for minority rule the spin-flip rate $w(S', S''\rightarrow
-S',-S'')\equiv w(S',S'')=W(S\rightarrow -S)$.

First consider majority rule dynamics.  In this case a given spin $S_j$
belongs to the three groups $(S_{j-2}, S_{j-1}, S_j)$, $(S_{j-1}, S_{j},
S_{j+1})$, and $(S_{j}, S_{j+1}, S_{j+2})$.  This then leads to the total
flip rate \cite{Majority}
\begin{eqnarray}
\label{sfMM1}
W(S_j\!\rightarrow\! -S_j) &\!=\!& 3 \!+\! S_{j-2} S_{j-1} \!+\! S_{j-1} S_{j+1} \!+\!
S_{j+1}S_{j+2} \nonumber\\
&\!-\!& S_{j} [2S_{j-1}\!+\!2S_{j+1}\!+\!S_{j-2}+S_{j+2}].
\end{eqnarray}
On the other hand, for minority rule, the spin-flip rates are 
\begin{eqnarray}
\label{sfMMM1}
 w(S_{j-2}, S_{j-1})
 &\!=\!& 1\!+\!S_{j-1}S_{j-2} \!-\! S_{j}(S_{j-1}+S_{j-2}), \nonumber\\
w(S_{j-1}, S_{j+1})
 &\!=\!& 1\!+\!S_{j-1}S_{j+1}\!-\!S_{j}(S_{j-1}+S_{j+1}), \nonumber\\
w(S_{j+1}, S_{j+2})
 &\!=\!& 1\!+\!S_{j+1}S_{j+2}\!-\!S_{j}(S_{j+1}+S_{j+2}).
\end{eqnarray}

The kinetics of the system is described by the master equation for the
probability distribution for a given spin configuration $\{S\}$.  The
derivation of this master equation is standard but tedious and the details
are given in Appendix A.  From the master equation, we can then compute the
rate equation for the magnetization (Eq.~(\ref{eqmagn})).  For the present
discussion, we only study a spatially homogeneous system.  In this case
Eq.~(\ref{eqmagn}) simplifies considerably and the resulting rate equation is
\begin{eqnarray}
\label{eqmagn1}
\frac{dm_1(t)}{dt} = 6\,(3p -2)\,(m_1(t) -m_3(t)),
\end{eqnarray}
with the magnetization $m_1(t)\equiv \langle S_{j} (t)\rangle$ written as the
first moment of the spin expectation value, and $m_{3}(t)\equiv \langle S_{j}
(t) S_{j+1} (t) S_{j+2} (t)\rangle$ is the 3-spin correlation function.

Notice that this equation has a very similar structure to Eq.~(\ref{magnMF}),
the mean-field equation for the magnetization.  In fact, Eq.~(\ref{eqmagn1})
reduces to (\ref{magnMF}) if we neglect fluctuations and assume that
$m_3=m_1^3$.  From Eq.~(\ref{eqmagn1}), we deduce several basic facts:

\begin{itemize}
\item For $p=p_c=\frac{2}{3}$ and $\forall~ m_1(0)$, the magnetization is
  conserved.  This conservation, valid in all spatial dimension, relies on
  the fact that the group size equals 3.  Thus at $p_c$ we expect kinetics
  similar to that in the classical voter model.
  
\item For any $p$, a system that is initially in consensus ($m_1(0)=\pm 1$)
  or a system with zero initial magnetization $(m_1(0)=0$) does not evolve.
  That is, $m_1(t)=m_1(0)=\pm 1$ in the former case and $m_1(t)=m_1(0)=0$ in

\item The magnetization is generally {\it not\/} conserved, except for the
  initial state $m_1(0)=0$ or $\pm 1$.  This non-conservation leads to
  unusual kinetics of the interfaces between regions of plus and minus spins.
  While these domain walls diffuse if they are widely separated, MM dynamics
  leads to additional interactions between walls when their distance is less
  than or equal to 2.
  
\item For $p\neq p_c$, the equation for the magnetization is not closed but
  involves the 3-spin correlation function.  In turn, the equation for this
  correlation function involves higher-order correlations, thus giving rise
  to an insoluble, infinite equation hierarchy.

\end{itemize}

To make analytical progress for the behavior of the magnetization in one
dimension, we need to truncate this equation hierarchy.  In the next section,
we implement such a truncation within the Kirkwood approximation scheme.

\subsection{Kirkwood approximation for the final magnetization}

We now study the behavior of the magnetization in one dimension.  Contrary to
the case of spatial dimension $d>1$, the magnetization quickly approaches a
saturation value that has a smooth and non-trivial dependence on $m_1(0)$
\cite{Majority}.  We implement a Kirkwood decoupling scheme to the exact
master equation to obtain the mean magnetization $m(t)$.  We shall see that
this uncontrolled approximation gives surprisingly accurate results.

Our approach is based on writing the exact equation of motion for
$m_2(t)=\langle S_j (t) S_{j+1} (t) \rangle$ and then, in the spirit of the
Kirkwood approximation \cite{K}, factorizing the 4-point functions that
appear in this equation as products of 2-point functions.  Such an approach
has proven quite successful in a variety of applications to reaction kinetics
\cite{SVB,lin,FK}.  By solving the resulting nonlinear but closed equation,
we obtain an approximate expression for $m_2$.  Then in Eq.~(\ref{eqmagn1})
for the magnetization, we factorize the 3-point function $m_3$ as $m_1m_2$
(instead of $m_1^3$ as in the usual mean-field analysis).

We now determine the equation of motion for the correlation function $m_2$
from the master equation (\ref{master}).  Following the same steps as those
followed to find the equation for the mean magnetization, we find, after a
number of straightforward steps (see Appendix~B),
\begin{eqnarray}
\label{c1mmm}
 \frac{dm_2}{dt} &=&4\Big[2(1+p) -(4+p) \, m_2 + p(c_2 + c_3)\Big]\nonumber\\
&+&4(2-3p) \, m_4,
\end{eqnarray}
where we have used the shorthand notations (for a translationally invariant
system) $c_{r}(t)\equiv \langle S_j (t) S_{j+|r|}(t) \rangle$ and
$m_4(t)\equiv \langle S_j (t) S_{j+1}(t)S_{j+2}(t)S_{j+3}(t)\rangle$.  In
general, we reserve the notation $m_{2k}$ to denote the average value of a
chain of $2k$ contiguous spins and $c_r$ for the correlation function between
two spins that are separated by a distance $r$.  Thus when the separation
between the two spins equals one, we have that $c_1(t)=m_2(t)\equiv \langle
S_j(t) S_{j\pm 1}(t) \rangle$.

In spite of the fact that Eq.~(\ref{c1mmm}) is exact, the 2-spin correlation
function $c_r(t)$ is coupled to higher-order correlations and it is therefore
difficult to compute these quantities exactly.  However at $p_c=
\frac{2}{3}$, this equation is closed in that it involves 2-spin correlation
functions only (see Sec.~III.C).  For $p\neq p_c$ we simply write $m_4$ as
$m_2^2$ in Eq.~(\ref{c1mmm}), following the Kirkwood approximation.  Since we
are mainly interested in the stationary state at $t=\infty$, where the
variation in the 2-point function as a function of $r$ is weak, we also make
the assumption that $c_2\approx c_3\approx m_2$.  

We show in Sec.~III.D that this approximation is accurate for the voter model
limit of $p=p_c$ and our numerical results also show that this approximation
continues to give a reasonable description for the properties of the final
state when $p\approx p_c$.  It is true, however, that this approach does not
provide a good description of the time dependence of the magnetization. 

With these approximations and for $p\neq \frac{2}{3}$, Eq.~(\ref{c1mmm})
becomes
\begin{eqnarray}
\label{c1appr}
 \frac{d m_2}{dt} =4(2-3p)\left[(m_2-1) \,
 \left(m_2 - \frac{2(1+p)}{2-3p}\right)\right].
\end{eqnarray}
Eq.~(\ref{c1appr}) admits $m_2 (\infty)=1$ as the unique and physically
acceptable fixed point.  (The other fixed points are $m_2^*=\frac{2(1+p)}{2-3p}>
1$ for $0<p<\frac{2}{3}$ and $m_2^*=\frac{2(1+p)}{2-3p}< -1$ for
$p>\frac{2}{3}$.)~ The general solution to Eq.~(\ref{c1appr}), for $0<p\leq
1$ and $p\neq p_c=\frac{2}{3}$, is
\begin{eqnarray}
\label{solm2I}
m_2(t)=\frac{A+\beta\,e^{-20 pt}}{A-e^{-20 pt}},
\end{eqnarray}
where 
\begin{eqnarray*}
\beta=\frac{p_c \, (1+p)}{p-p_c}\qquad {\rm and}\quad A=\frac{m_2(0)+\beta}{m_2(0)-1}
\end{eqnarray*}
At $p_c=\frac{2}{3}$, we obtain $m_2(t)= 1-[1-m_2(0)]e^{-40t/3}$.  Thus, 
for all $p$, $m_2(t)\to 1$ as $t\to\infty$.

We now exploit this result to compute the final magnetization.  In the exact
equation (\ref{eqmagn1}) for $m_1$, we write $m_3$ as $m_1m_2$ to give 
\begin{eqnarray}
\label{eqmagnKirk}
\frac{dm_1}{dt} =6(3p-2) \, {m_1} \,(1-m_2).
\end{eqnarray}
Notice a crucial difference between this equation of motion and the
mean-field equation (\ref{magnMF}).  In the stationary state,
Eq.~(\ref{eqmagnKirk}) predicts that either $m_1(\infty)=0$ or
$m_2(\infty)=1$.  Since $m_2(t)\neq m_1(t)^2$ in the Kirkwood approximation,
this means that $m_1(\infty)$ can be a non-trivial function, even if
$m_2(\infty)=1$.

\begin{figure}[ht] 
 \vspace*{0.cm}
 \includegraphics*[width=0.4\textwidth]{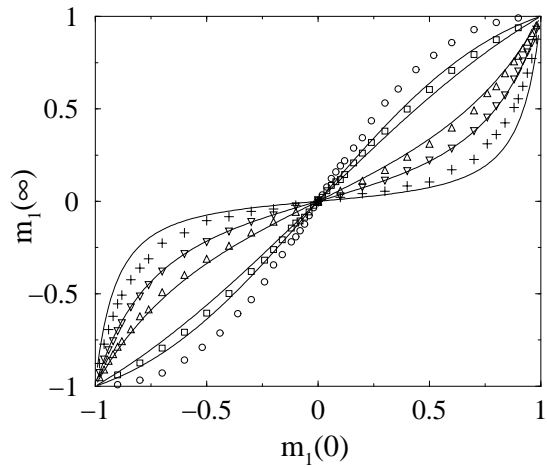}
\caption{The final magnetization as a function of the initial magnetization.
  Shown are the results of numerical simulations for the cases $p=0.1$ ($+$),
  0.25 ($\nabla$), 0.4 ($\Delta$), 0.8 ($\square$) and 1 ($\circ$).  The
  smooth curves are the corresponding results from the
  Kirkwood approximation (Eq.~(\ref{magnKirkbis})).}
\label{m-final}
\end{figure}

Integrating Eq.~(\ref{eqmagnKirk}) gives the formal expression for the final
magnetization
\begin{eqnarray}
\label{magnKirk0}
m_1(\infty) \!=\! m_1(0)\exp\Big\{18(p\!-\!p_c) 
\int_{0}^{\infty}\!\!\!\! dt' [1\!-\!m_2(t')] \Big\}.
\end{eqnarray}
Substituting the expression for $m_2(t)$ in Eq.~(\ref{solm2I}), we thereby
obtain
\begin{eqnarray}
\label{magnKirk}
m_1(\infty) = m_1(0) \; \left(\frac{\beta+1}{\beta + m_2(0)}\right)^{3/2} .
\end{eqnarray}
For an initially uncorrelated and random system, $m_{2}(0)=m_{1}(0)^{2}$, and
\begin{eqnarray}
\label{magnKirkbis}
m_1(\infty) = m_1(0) \; \left(\frac{\beta+1}{\beta + m_1(0)^2}\right)^{3/2}.
\end{eqnarray}
Thus the Kirkwood approximation predicts a final magnetization that is a
non-trivial function of the initial magnetization (Fig.~\ref{m-final}).  As
$p \to p_c=\frac{2}{3}$ this approximation correctly predicts that the
average magnetization is conserved, that is, $m_1(t) = m_1(0)$.  When $p\to
0$, this approximation also predicts (for $m_1(0)\neq \pm 1$) that the final
magnetization vanishes ({\it i.e.}  $m_1(\infty) = 0$).  Fig.~\ref{m-final}
shows that the Kirkwood approximation is quantitatively accurate for
intermediate values of $p$ but is only qualitative for $p$ close to either 0
or 1.

\subsection{2-spin correlation function at $p_c$ }

At $p_c=\frac{2}{3}$, Eq.~(\ref{eqmagn1}) shows that the magnetization of the
MM is conserved.  This same conservation law occurs in the voter model which
has a consequence that the correlation function $\langle
S_{j}(t)S_{j+r}(t)\rangle-1$ vanishes as $t^{-1/2}$ in one dimension.  We now
show that this same type of coarsening also occurs in the MM by computing the
2-spin correlation functions at $p_c$.  The equations of motion for these
correlation functions are cumbersome and they are written in Appendix B.

For our purposes, we concentrate on translationally invariant and symmetric
systems.  Then in the equations of motion for the 2-point function
[Eqs.~(\ref{corrMM})--(\ref{corrMMr2})], the coordinates $j_1, j_2, j_3, j_4$
in the 4-point functions, always appear as three consecutive positions, then
a gap of size $r$, followed by the coordinate of the last spin.  This gap can
either occur on the left or the right side of the spin group.  To simplify
the notation, we therefore write these 4-point ``gap'' functions of the form
$\langle S_{j-2}(t)S_{j-1}(t)S_{j}(t)S_{j+r}(t)\rangle$ and $\langle
S_{j}(t)S_{j+r}(t)S_{j+r+1}(t)S_{j+r+2}(t)\rangle$ as ${\cal G}_r(t)$.  With
these simplifications Eq.~(\ref{corrMM}) gives, for the case of majority rule
({\it i.e.}, $p=1$) and for $|r|>2$
\begin{eqnarray}
\label{corrMM_TI}
\frac{dc_r}{dt}  &=&
4\left(c_{r+2} +c_{r-2} +2c_{r+1}+2c_{r-1} -3c_r\right)
\nonumber\\
&-&4\left({\cal G}_r+ {\cal G}_{r-1}+{\cal G}_{r-2}\right).
\end{eqnarray}
For $r=1$, Eq.~(\ref{corrMMr1}) gives
\begin{eqnarray}
\label{corrMMr1_TI}
\frac{dc_1}{dt}  &=& 4\left(c_{2} +c_{3}+4 - 5c_1 - {\cal G}_1 \right),
\end{eqnarray}
while Eq.~(\ref{corrMMr2}) gives
\begin{eqnarray}
\label{corrMMr2_TI}
\frac{dc_2}{dt}  &=&
4\left(2+c_1 +2c_3 + c_4- 4c_2  \right) \nonumber\\&-&
4\left({\cal G}_1+ {\cal G}_2  \right)
\end{eqnarray}
Together with $c_0 (t) =1$, Eqs.~(\ref{corrMM_TI})--(\ref{corrMMr2_TI})
are the equations of motion for $c_r(t)$ for the translationally-invariant
majority model.

For the minority model ($p=0$) we proceed in a similar manner to write the
analog of Eq.~(\ref{corrMM}).  After straightforward but lengthy computations
the equation of motion is (for $|r|>2$)
\begin{eqnarray}
\label{corrMinor_TI}
\frac{dc_r}{dt}  &=&
8\left({\cal G}_r+ {\cal G}_{r-1}+{\cal G}_{r-2}\right)-24 c_r,
\end{eqnarray}
while for $r=1$
\begin{eqnarray}
\label{corr_r1_Minor_TI}
\frac{dc_1}{dt}  = 8(1-2c_1+{\cal G}_1),
\end{eqnarray}
and for $r=2$ we have
\begin{eqnarray}
\label{corr_r2_Minor_TI}
\frac{dc_2}{dt}  = 8\left(1 - 3c_2 +{\cal G}_1 +{\cal G}_2 \right).
\end{eqnarray}
These equations again have to be supplemented by the boundary condition
$c_0(t)=1$.

The equation of motion for the 2-spin correlation function in the MM model
can now be obtained by taking $p$ times Eq.~(\ref{corrMM_TI}) and $1-p$ times
(\ref{corrMinor_TI}).  For general $p$, this leads to an open equation
hierarchy.  However, at $p=p_c=\frac{2}{3}$, the 4-spin correlation functions
arising from both the majority and minority models {\it cancel\/} for all
values of $r$.  Thus at $p=p_c$, we obtain much simpler equations for motion
for the 2-spin correlation function.  For $|r|>2$, we obtain
\begin{equation}
\label{crMMM}
\frac{dc_r}{dt}  =
\frac{8}{3}\left[
c_{r+2} +c_{r-2} +2(c_{r+1} +c_{r-1})-6c_r\right].
\end{equation}
For $r=1$ we obtained previously (Eq.~(\ref{c1mmm}))
\begin{equation}
\label{c1MMM}
\frac{dc_1}{dt} =
\frac{8}{3}\left(5+c_{2}+c_{3}-7c_1\right),
\end{equation}
while for $r=2$ we have 
\begin{equation}
\label{c2MMM}
\frac{dc_2}{dt} =\frac{8}{3}\left(3+c_{4}+2c_{3} +c_{1}-7c_2  \right).
\end{equation}
Eqs.~(\ref{crMMM})--(\ref{c2MMM}), together with the boundary condition
$c_0(t)=1$, constitute a closed and soluble set of coupled linear
differential-difference equations for the 2-spin correlation functions.

\subsection{Solution for the 2-spin correlation function}

To solve Eqs.~(\ref{crMMM})--(\ref{c2MMM}), first notice that these coupled
equations can be recast as the single equation for the auxiliary quantity
$g_r(t)\equiv c_r (t)-1$
\begin{eqnarray}
\label{greq}
\frac{dg_r}{dt}  &=&
g_{r+2}+g_{r-2}+2(g_{r+1}+g_{r-1})-6g_r\nonumber\\ 
&-& (g_2+ 2g_1) (\delta_{r,1}+\delta_{r,-1})- (g_1+g_2)
(\delta_{r,2}+\delta_{r,-2})\nonumber \\
&-& 2(g_2+2g_1)\delta_{r,0},
\end{eqnarray}
where for simplicity we have also rescaled the time according to
$t\rightarrow \frac{8}{3}t$.

Before proceeding, it is instructive to recall that in the one-dimensional
voter model, the equation for the 2-spin correlation function $c_{r}^{vm}(t)$
for a translationally-invariant system has the form of the discrete diffusion
equation \cite{glauber}
\begin{eqnarray}
\label{crVMeq}
\frac{d}{dt}c_r^{vm}  &=&
c_{r+1}^{vm}+c_{r-1}^{vm}-2c_r^{vm}
\end{eqnarray}
for $|r|\geq 1$, supplemented by the boundary condition $c_{0}^{vm}(t)=1$.
The solution to this equation is
\begin{eqnarray}
\label{crVM}
c_r^{vm} (t) \!=\! 1\!+\!e^{-2t}\sum_{l=1}^{\infty}[c_{l}^{vm}(0)\!-\!1]
\left[I_{r\!-\!l}(2t)\!-\! I_{r\!+\!l}(2t) \right],
\end{eqnarray}
where $I_{r}(t)$ is the modified Bessel function of first kind
\cite{Abramowitz}.

For the MM model, Eq.~(\ref{greq}) is also a discrete diffusion equation but
with second-neighbor hopping.  Thus we expect that this equation can be
solved by similar techniques as those used in the voter model.  Therefore we
introduce the following integral representation that generalizes the modified
Bessel function of the first kind
\begin{eqnarray}
\label{I}
{\cal I}_r(t)\equiv \frac{1}{\pi}\int_{0}^{\pi}dq \; \cos{(qr)} \,
e^{2t[\cos{2q} + 2\cos{q}]}.
\end{eqnarray}
It is easy to check that ${\cal I}_r(t)$ satisfies the basic recursion
property $\dot {\cal I}_r(t)={\cal I}_{r-2}(t) +{\cal I}_{r+2}(t) + 2\left[
  {\cal I}_{r-1}(t) +{\cal I}_{r+1}(t) \right]$.  Also in analogy with the
modified Bessel function of first kind, $\sum_{r=-\infty}^{+\infty}{\cal
  I}_r(t)=e^{6t}$ and ${\cal I}_r (0)=\delta_{r,0}$.  With these properties,
the formal solution of Eq.~(\ref{greq}) is
\begin{align}
\label{grsol}
g_r (t)&=e^{-6t}\sum_{r'=-\infty}^{+\infty} g_{r'}(0) \,{\cal I}_{r-r'}(t)\nonumber\\
- \int_{0}^{t} \!dt' &g_1(t\!-\!t')\, e^{-6t'}\,
\Big[4 {\cal I}_{r}(t')\!+\!2\sum_{r_1}{\cal I}_{r_1}(t')\!+\!\sum_{r_2}{\cal
  I}_{r_2}(t')\Big] \nonumber \\
- \int_{0}^{t} \!dt' &g_2(t\!-\!t')\, e^{-6t'}\,
 \Big[2 {\cal I}_{r}(t')\!+\!\sum_{r_{1,2}}{\cal I}_{r_{1,2}}(t')\Big], 
\end{align}
where in the second line the sums are over the nearest- and next-nearest
neighbors of $r$ respectively, while in the third line the sum is over both
nearest- and next-nearest neighbors.

Since the right-hand side of (\ref{grsol}) still depends on $g_1$ and $g_2$,
we have to consider the cases $r=1$ and $2$ separately to obtain the general
solution. This is done in Appendix C by using Laplace transforms.  In the
long-time and large-distance limit, the full solution to (\ref{grsol}) quoted
in Eq.~(\ref{cR(t)long'}) reduces to the much simpler expression
\begin{eqnarray}
\label{cR(t)long''}
 c_r(t)\simeq m_1(0)^2+
\left[1-m_1(0)^2\right] \, {\rm erfc}\left(\frac{r}{8\sqrt{t}}\right)
\end{eqnarray}
that clearly shows the scaling behavior in $r$ and $t$.  For comparison, the
2-spin correlation function of the voter model, in the same limit and with
the same initial condition of $c_{r}^{vm}(0)=m_0^2$, is
\begin{eqnarray}
\label{cR(t)VMlong}
 c_r^{vm}(t)\simeq m_0^2- \frac{(1-m_0^2)}{2\sqrt{\pi t }} 
\sum_{1\leq l \leq 2r} e^{-\frac{(r-l)^2}{4t}}.  
\end{eqnarray}

Comparing these two results, we see that the MM model shares many of the
asymptotic features of the voter model.  The correlation between spins that
are separated by a fixed distance $r$ both approach the value one, with the
deviation from the asymptotic value decaying as $t^{-1/2}$.  As in the voter
model, the density of domain walls between regions of plus and minus spins,
that is, $(1-c_1(t))/2$ decays as $t^{-1/2}$ [see Eq.~(\ref{c(t)long''})].
 Thus in the one-dimensional MM
there is coarsening with typical domains growing as $t^{1/2}$, as in the
voter model \cite{LFPK}.

Our exact results also sheds light on the basic nature of the Kirkwood
approximation.  This approximation gave $c_{1}(t)= 1-[1-m_1(0)^2]e^{-40t/3}$,
whereas the exact result of Eq.~(\ref{c(t)long''}) predicts that $c_1(t)$
approaches one with a correction term proportional to $t^{-1/2}$.  Although
both expressions give the same asymptotic state of consensus, the incorrect
time dependence in the Kirkwood approximation appears to stem from our
assumption that $c_1(t)\approx c_2(t)\approx c_3(t)$.  Although this is valid
in the stationary state, it is certainly incorrect in the transient regime
where this assumption is at odds with the diffusive nature of the problem.
As confirmed by numerical results, we thus expect that the Kirkwood
approximation should give good results for the stationary magnetization, but
not for the approach to this state.

\section{Summary and Discussion}

We introduced a simple model of opinion dynamics -- termed the MM model -- in
which a fixed-size group of agents is specified and all members of the group
adopt the local majority state with probability $p$ or the local minority
state with probability $1-p$.  We considered the simplest case where the
group size $G=3$.  In the mean-field limit, the probability that the system
ends with all spins plus as a function of the initial magnetization of the
system (the exit probability) can be readily obtained.  For
$p>p_c=\frac{2}{3}$, this exit probability changes abruptly from $-1$ for
initial magnetization $m(0)<0$ to $+1$ for $m(0)>0$.  Conversely, for
$p<p_c$, this exit probability is 1/2 for almost all $m(0)$.  These behaviors
reflect the inherent biases of majority and minority rule.

In one dimension, the magnetization quickly approaches a fixed value that
depends only on the initial magnetization.  This then immediately determines
the exit probability.  Within a Kirkwood decoupling scheme for the infinite
hierarchy of equations for correlation functions, we obtained a reasonable
approximation for the dependence of the final magnetization (equivalently the
exit probability) on the initial magnetization.  It is worth noting that
other decoupling schemes can also be applied.  One such example is the
so-called ``simple method'' \cite{simple}, where the 3- and 4-point
correlation functions are decoupled according to $m_3=m_2^2/m_1$ and
$m_4=m_2^2$.  While this approach sometimes gives superior results to the
Kirkwood scheme \cite{lin}, this approach turns out to be ill suited to
determining the initial density dependence of the final magnetization in one
dimension.

At the critical point of $p_c=\frac{2}{3}$, we obtained the exact 2-spin
correlation function and showed that it exhibits the same $t^{1/2}$
coarsening as in the classical voter model.  Although the 2-spin correlation
function has the same behavior as in the voter model, it is possible that
two-time correlation functions, such as $\langle S(t)S(t')\rangle$, or
quantities related to persistence phenomena, will give behavior different
that the voter model.

We would like to suggest several directions for further research.  First, it
would be worthwhile to understand the MM model in finite spatial dimensions
strictly greater than one.  In the special case where the majority
exclusively rules ($p=1$), numerical evidence suggested that the upper
critical dimension of the system is greater than four \cite{Majority}.  On
the other hand, the upper critical dimension for the voter model equals two
and this appears to coincide with the behavior of the MM model for $p=p_c$.
It should be instructive to understand the nature of the crossover between
these two behaviors.

Another question involves the dependence of the kinetics on the group size.
For group size $G>3$, a sharp transition between majority-dominated and
minority-dominated kinetics can be engineered by the following somewhat
baroque construction.  For a group that contains $k$ plus spins and $G-k$
minus spins, apply majority rule with probability $k/G$ and minority rule
with probability $1-k/G$.  It is easy to verify that this rule gives zero net
magnetization change in each elemental group update.  Thus this construction
should lead to kinetics similar to that of the voter model.  However, in the
more natural situation where the probabilities of applying the majority or
minority rules are independent of group composition, we do not yet understand
the nature of the change between majority-dominated and minority-dominated
dynamics.

The kinetics in one dimension presents an intriguing challenge.  Within the
Glauber formalism, the MM model appears to be insoluble because correlation
functions of different orders are coupled in the equations of motion.
However, the evolution of interfaces between domain walls obeys relatively
simple kinetics that closely resembles the diffusion-limited reaction $A+A\to
0$.  For the MM model, it is easy to see that, in addition to diffusion of
domain walls, there are specific constraints in their motion when domain
walls are either nearest-neighbor or next-nearest-neighbor.  In spite of
these complications, we would hope that this model is exactly soluble in one
dimension.

Finally, it should be worthwhile to extend the model to allow for agents that
have an intrinsic identity.  In the MM model, the state of an agent is
determined only by the local environment.  However, it is much more realistic
for individuals to inherently prefer one of the two states so that the
transition rates depend both on this factor, as well as on the state of its
neighbors.  This seems a natural step to bring the MM model a bit closer to
political reality.

\section{Acknowledgements}

We thank Paul Krapivsky for many helpful discussions and advice.  M.M. thanks
the Swiss NSF, under the fellowship 81EL-68473, and S.R. thanks NSF grant
DMR0227670 for financial support of this research.

\begin{widetext}
\appendix

\section{Master equation}

We write the master equation for the probability distribution of a given spin
configuration and then use this to obtain the equation of motion for the
magnetization.  From the definition of the MM, the master equation is
\begin{eqnarray}
\label{master}
\frac{d}{dt}P(\{S\},t)&=& p\sum_{k}\left[
W(-S_k\rightarrow S_k)P(\{S\}_k,t) -
W(S_k\rightarrow -S_k)P(\{S\},t) \right] \nonumber\\
&+& (1-p) \sum_{k}\left[
w(-S_{k-2}; -S_{k-1})P(\{S\}_{k-2,k-1},t) -
w(S_{k-2}; S_{k-1})P(\{S\},t) \right] \nonumber\\
&+& (1-p) \sum_{k}\left[
w(-S_{k-1}; -S_{k+1})P(\{S\}_{k-1,k+1},t) -
w(S_{k-1}; S_{k+1})P(\{S\},t) \right] \nonumber\\
&+& (1-p) \sum_{k}\left[
w(-S_{k+1}; -S_{k+2})P(\{S\}_{k+1,k+2},t) -
w(S_{k+1}; S_{k+2})P(\{S\},t) \right]. \nonumber\\
\end{eqnarray}
Here $P(\{S\}, t)$ denotes the probability for the spin configuration $\{S\}$
at time $t$ and $P(\{S\}_k, t)$ is the probability for the configuration
$\{S\}_k$ where spin $S_k$ is reversed compared to $\{S\}$.  Similarly
$P(\{S\}_{k_1,k_2}, t)$, is the probability of the configuration where spins
$S_{k_1}$ and $S_{k_2}$ are reversed compared to $\{S\}$.

From this master equation, and with help of Eqs.~(\ref{sfMM1}) and
(\ref{sfMMM1}), it follows that the mean magnetization obeys the equation of
motion
\begin{eqnarray}
\label{eqmagn}
\frac{d}{dt}\langle S_j \rangle&=&\sum_{\{S \}} S_j
\frac{d}{dt}P(\{S\},t) \nonumber \\
&=&2p \, [ \langle S_{j-2}\rangle + \langle S_{j+2}\rangle
+2 \langle S_{j-1} \rangle + 2 \langle S_{j+1} \rangle -3\langle S_j\rangle]\nonumber \\
&&-2p\left[\langle S_j S_{j+1} S_{j+2} \rangle + \langle S_{j-1} S_j S_{j+1}
\rangle +  \langle S_{j-2} S_{j-1} S_{j} \rangle\right]\nonumber\\
&&-2(1-p)\left[6\langle S_j\rangle
- 2\langle S_{j-2} S_{j-1} S_j\rangle - 2\langle S_{j-1} S_j S_{j+1}\rangle - 
2\langle S_j S_{j+1} S_{j+2} \rangle\right]
\end{eqnarray}
To arrive at this equation, we have taken the thermodynamic limit, made some
obvious cancellations, and used the following relations:
\begin{eqnarray*}
\sum_{\{S\}} S_{j}P(\{S\}_{j})= -\langle S_j \rangle &;& \quad
\sum_{\{S\}} S_{j}P(\{S\}_{k\neq j})= \langle S_j \rangle \nonumber\\
\sum_{\{S\}} S_{j}P(\{S\}_{j, k'\neq j})= -\langle S_j \rangle &;& \quad
\sum_{\{S\}} S_{j}P(\{S\}_{k\neq j, k'\neq j})= \langle S_j \rangle \nonumber\\
\sum_{\{S\}} S_{j}S_{j'}P(\{S\}_{j,j'})= \langle S_j S_{j'} \rangle
\end{eqnarray*}

\section{Equations for motion for 2-spin correlation functions}

We write the general equations of motion for the 2-spin correlation
functions.  For simplicity consider the case of majority rule ({\it i.e.},
$p=1$).  In this case, we have
\begin{eqnarray}
\label{corrMM}
&&\frac{d}{dt}\langle S_j(t) S_{j+r}(t)\rangle =
-2\langle S_j \,  S_{j+r} \, W(S_j \rightarrow -S_j) \rangle
-2\langle S_j \,S_{j+r}\, W(S_{j+r} \rightarrow -S_{j+r}) \rangle
\nonumber\\
&=&2\left[
\langle S_{j-2}(t) S_{j+r}(t)\rangle +  \langle S_{j+2}(t) S_{j+r}(t)\rangle
+2\langle S_{j-1}(t) S_{j+r}(t)\rangle+ 2\langle S_{j+1}(t) S_{j+r}(t)\rangle
-6\langle S_{j}(t) S_{j+r}(t)\right]\rangle \nonumber\\
&+& 2\left[\langle S_{j}(t) S_{j+r-2}(t)\rangle
 + \langle S_{j}(t) S_{j+r+2}(t)\rangle +  2\langle S_{j}(t)
 S_{j+r-1}(t)\rangle
+ 2\langle S_{j}(t) S_{j+r+1}(t)\rangle
\right] \nonumber\\
&-&2\left[
\langle S_{j-2}(t) S_{j-1}(t)S_{j}(t) S_{j+r}(t)\rangle
+ \langle S_{j-1}(t) S_{j}(t)S_{j+1}(t) S_{j+r}(t)\rangle
+\langle S_{j}(t) S_{j+1}(t)S_{j+2}(t) S_{j+r}(t)\rangle
\right] \nonumber\\&-&
2\left[
\langle S_{j}(t) S_{j+r-2}(t)S_{j+r-1}(t) S_{j+r}(t)\rangle
+ \langle S_{j}(t) S_{j+r-1}(t)S_{j+r}(t) S_{j+r+1}(t)\rangle
+\langle S_{j}(t) S_{j+r}(t)S_{j+r+1}(t) S_{j+r+2}(t)\rangle
\right]. \nonumber\\
\end{eqnarray}
This equation applies for $r\neq 0, \pm1, \pm2$.  For $r=0$ we have simply
$\langle S_{j}(t)^2 \rangle=1$.  The cases $r=\pm 1$ and $r=\pm2$ have to be
dealt with separately.  For $r=1$, we have
\begin{eqnarray}
\label{corrMMr1}
\frac{d}{dt}\langle S_j(t) S_{j+1}(t)\rangle &=&
2\left[8+\langle S_{j-2}(t) S_{j+1}(t)\rangle
-10\langle S_{j}(t) S_{j+1}(t) \rangle +
\langle S_j(t) S_{j+2}(t)\right.\rangle\nonumber \\
&+&\left.\langle S_{j-1}(t) S_{j+1}(t)\rangle
+\langle S_{j}(t) S_{j+3}(t) \rangle\right]
\nonumber\\
&-& 2\left[\langle S_{j-2}(t)S_{j-1}(t)S_{j}(t)S_{j+1}(t)\rangle
+\langle S_{j}(t)S_{j+1}(t)S_{j+2}(t)S_{j+3}(t)\rangle
 \right]
\end{eqnarray}
and the equation $r=-1$ has a very similar form.
For $r=2$ we obtain
\begin{eqnarray}
\label{corrMMr2}
\frac{d}{dt}\langle S_j(t) S_{j+2}(t)\rangle &=&
2\left[
4-8\langle S_{j}(t) S_{j+2}(t) \rangle
+2 \langle S_{j}(t) S_{j+3}(t) \rangle
+ 2 \langle S_{j-1}(t) S_{j+2}(t) \rangle
+ \langle S_{j}(t) S_{j+4}(t) \rangle \right]
\nonumber\\
&+& 2 \left[\langle S_{j-2}(t) S_{j+2}(t) \rangle
+\langle S_{j}(t) S_{j+1}(t) \rangle
+ \langle S_{j+1}(t) S_{j+2}(t) \rangle
\right] \nonumber\\
&-&2\left[
\langle S_{j-2}(t) S_{j-1}(t)S_{j}(t) S_{j+2}(t)\rangle
+ \langle S_{j-1}(t) S_{j}(t)S_{j+1}(t) S_{j+2}(t)\rangle
\right] \nonumber\\&-&
2\left[
 \langle S_{j}(t) S_{j+1}(t)S_{j+2}(t) S_{j+3}(t)\rangle
+\langle S_{j}(t) S_{j+2}(t)S_{j+3}(t) S_{j+4}(t)\rangle
\right],
\end{eqnarray}
and similarly for $r=-2$.  For a translationally invariant system,
Eqs.~(\ref{corrMM})--(\ref{corrMMr2}) reduce respectively to
Eqs.~(\ref{corrMM_TI})--(\ref{corrMMr2_TI}).

The equations of motion for minority rule (where $p=0$) are obtained in a
similar manner by starting with the analog of Eq.~(\ref{corrMM}) when the
minority rule hopping rates are used.  For the translationally invariant
minority model, the equations of motion for the correlation functions are
then given by Eqs.~(\ref{corrMinor_TI})--(\ref{corr_r2_Minor_TI}).

\section{solution for the correlation function}

In this appendix, we solve Eq.~(\ref{grsol}).  For this purpose it is
convenient to introduce the Laplace transform.  For an uncorrelated but
random initial state where $c_r(0)=m_1(0)^2$, and using the properties of the
functions ${\cal I}$ introduced in Eq.~(\ref{I}), the Laplace transform of
$g_r(t)$ is
\begin{eqnarray}
\label{gLaplT}
\hat{g}_r (s)&\equiv& \int_{0}^{\infty} dt\, e^{-st}\,
g_r(t) \nonumber\\&=&
-\frac{1-[m_1(0)]^2}{s}- [
4\hat{{\cal I}}_{r}(s) 
+2\hat{{\cal I}}_{r+1}(s)+2\hat{{\cal I}}_{r-1}(s)+
 \hat{{\cal I}}_{r+2}(s) + \hat{{\cal I}}_{r-2}(s)
 ]\hat{g}_1 (s)
\nonumber\\
& &- [2\hat{{\cal I}}_{r}(s)+\hat{{\cal I}}_{r+1}(s)+\hat{{\cal I}}_{r-1}(s)
+\hat{{\cal I}}_{r+2}(s)+\hat{{\cal I}}_{r-2}(s)
]\hat{g}_2 (s),
\end{eqnarray}
where
\begin{eqnarray}
\label{Ir(s)}
\hat{{\cal I}}_{r}(s) &\equiv& \int_{0}^{\infty} dt\, e^{-st} \, [e^{-6t}{\cal
  I}_{r}(t)]\nonumber \\
&=&\int_{0}^{\pi} \frac{dq}{\pi} \, \frac{\cos{qr}}{s+6-2\{\cos{2q}
  +2 \cos{q}\}}\nonumber\\&=&
i\oint_{\Gamma} \frac{dz}{2\pi} \; \frac{z^{r+1}}{z^4 +2z^3-(s+6)z^2 +2 z+1},
\end{eqnarray}
and $\Gamma$ denotes the unit circle in the complex plane centered at the
origin.  In principle, the integral (\ref{Ir(s)}) can be computed by the
residue theorem.  However, we shall see that this calculation is unnecessary
for determining the long-time behavior of the correlation functions.

By substituting $r=1$ and $r=2$ into (\ref{gLaplT}) we obtain a linear system
of two equations that is readily solved and gives, for the Laplace transforms
of $g_1(t)$ and $g_2(t)$,
\begin{eqnarray}
\label{g1(s)}
\hat{g}_1(s)&=&(1-[m_1(0)]^2) \, \frac{{\cal J}_2(s) - {\cal K}_2(s) -1}
{[1+ {\cal J}_1(s) +{\cal K}_2(s) -{\cal J}_2(s) {\cal K}_1(s) + 
{\cal J}_1(s) {\cal K}_2(s)]s}\nonumber \\
\hat{g}_2(s)&=&(1-[m_1(0)]^2) \, \frac{{\cal K}_1(s) - {\cal J}_1(s) -1}
{[1+ {\cal J}_1(s) +{\cal K}_2(s) -{\cal J}_2(s) {\cal K}_1(s) + 
{\cal J}_1(s) {\cal K}_2(s)]s},
\end{eqnarray}
where we have introduced the following quantities:
\begin{eqnarray}
\label{J1}
{\cal J}_1(s)&\equiv& 2\hat{{\cal I}}_0(s)+ 5 \hat{{\cal I}}_1(s)+
2\hat{{\cal I}}_2(s) + \hat{{\cal I}}_3(s)\nonumber \\
{\cal K}_1(s)&\equiv& 
\hat{{\cal I}}_0(s)+3\hat{{\cal I}}_1(s)+
 \hat{{\cal I}}_2(s)+\hat{{\cal I}}_3(s)\nonumber \\
{\cal J}_2(s)&\equiv& 
\hat{{\cal I}}_0(s)+ 2\hat{{\cal I}}_1(s) + 4\hat{{\cal I}}_2(s)+
2 \hat{{\cal    I}}_3(s) + \hat{{\cal I}}_4(s)\nonumber \\
{\cal K}_2(s)&\equiv&  \hat{{\cal I}}_0(s) + \hat{{\cal I}}_1(s) + 
2\hat{{\cal I}}_2(s)+\hat{{\cal I}}_3(s)+\hat{{\cal I}}_4(s)
\end{eqnarray}

Since we are mainly interested in the long-time behavior of the 2-spin
correlation functions, we focus on the small-$s$ dependence of the quantities
in (\ref{J1}).  For $s\rightarrow 0$ the integral (\ref{Ir(s)}) diverges for
$q\rightarrow 0$.  Clearly, the main contribution to this integral in the
long-time limit (equivalently $s\rightarrow 0$) is obtained by expanding the
integrand for $q\rightarrow 0$ before performing the integration.  We obtain
\begin{eqnarray}
\label{Irlarge}
\hat{{\cal I}}_r(s) \xrightarrow[s\rightarrow 0]{} \int_{0}^{\pi}
\frac{dq}{\pi}\, \frac{\cos{(qr)}}{s+6q^2}  
\simeq \int_{0}^{\infty}
\frac{dq}{\pi}\, \frac{\cos{(qr)}}{s+6q^2}=
\frac{e^{-r\sqrt{s/6}}}{2 \sqrt{6s}}.
\end{eqnarray}

Substituting this expression into Eq.~(\ref{g1(s)}) and expanding the
resulting exponential terms gives
\begin{eqnarray}
\label{g(s)'}
 \hat{g}_1(s)&\xrightarrow[s\rightarrow 0]{}&- \frac{2}{25}\, [1-m_1(0)^2]\,
 \sqrt{\frac{6}{s}}\nonumber \\
 \hat{g}_2(s)&\xrightarrow[s\rightarrow 0]{}&-\frac{1-m_1(0)^2}{5} \, \sqrt{\frac{6}{s}}
\end{eqnarray}
The expression of $\hat{g}_1$ and $\hat{g}_2$, together with (\ref{gLaplT}), provide the Laplace
transform of $\hat g_r(r)$ in the $s\rightarrow 0$ regime.

For finite $r>2$, we substitute (\ref{Irlarge}) and (\ref{g(s)'})  into (\ref{gLaplT}), expand
the exponential terms as $r\sqrt{s}\rightarrow 0$ and obtain
\begin{eqnarray}
\label{gr(s)long}
 \hat{g}_r(s)\xrightarrow[s\rightarrow 0]{}
- \left(1-m_1(0)^2\right) \;\frac{5r-4}{5\sqrt{6s}}.
\end{eqnarray}
Laplace inverting Eqs.~(\ref{g(s)'}) and
(\ref{gr(s)long}) then gives, for $t\rightarrow \infty$, with $r^2 \ll t$,
\begin{eqnarray}
\label{c(t)long''}
 c_1(t)&=&
1- \frac{3\, \left[1-m_1(0)^2\right] }{25\, \sqrt{\pi t}},  \nonumber \\
 c_2(t)&=&
1- \frac{3\, \left[1-m_1(0)^2\right] }{10\, \sqrt{\pi t}}, \nonumber \\
 c_r(t)&=&
1- \left[1-m_1(0)^2\right] \;\frac{5r-4}{20\sqrt{\pi t}}\hspace{2cm} (r\geq 2),
\end{eqnarray}
where we have restored the original time scale, {\it i.e.}, $t\to \frac{3}{8}t$.

In the limit $r\rightarrow \infty$ and $s\rightarrow 0$, with $r\sqrt{s}$
kept fixed, we substitute Eqs.~(\ref{g(s)'})  into
(\ref{gLaplT}), and obtain, after inverse Laplace transforming,
\begin{eqnarray}
\label{cR(t)long'}
 c_r(t)=[m_1(0)]^2 &+&
\frac{1-[m_1(0)]^2}{50}\left[
18 \,{\rm erfc}\left(\frac{r}{8\sqrt{t}}\right)+ 
7\,{\rm erfc}\left(\frac{r+2}{8\sqrt{t}}\right)
+ 7\,{\rm erfc}\left(\frac{r-2}{8\sqrt{t}}\right)
\right] \nonumber\\
&+& \frac{9\,(1-[m_1(0)]^2)}{50} \, \left[ 
{\rm erfc}\left(\frac{r+1}{8\sqrt{t}}\right)
+ {\rm erfc}\left(\frac{r-1}{8\sqrt{t}}\right)
\right].
\end{eqnarray}
where ${\rm erfc}(t)\equiv \frac{2}{\sqrt{\pi}} \int_{t}^{\infty} dz \,
e^{-z^2}$ is the complementary error function, and we used the fact that the
inverse Laplace transform of $e^{-\sqrt{sa}}/{s}$ is ${\rm
  erfc}\left(\frac{1}{2} \sqrt{\frac{a}{t}}\right)$ \cite{Abramowitz}.
Eq.~(\ref{cR(t)long'}) simplifies considerably if we make the $r\rightarrow
\infty$ approximation $r\approx r\pm 1\approx r\pm 2$.  In this limit, we
obtain the expression quoted in Eq.~(\ref{cR(t)long''}).

\newpage
\end{widetext}

\end{document}